\documentclass[twocolumn,tighten,times]{aastex62}
\usepackage{rotating}
\newcommand{\err}[2]{\ensuremath{^{_{+#1}}_{^{-#2}}}}
\newcommand{\ee}[2]{\ensuremath{{#1}\!\times\!10^{#2}}}
\newcommand{\msun}{\ensuremath{\mathrm{M}_\odot}}

\newcommand{\ergcms}{\ensuremath{\mathrm{erg~cm}^{-2}~\mathrm{s}^{-1}}}

\newcommand{\Fx}{\ensuremath{F_\mathrm{x}}}
\newcommand{\nH}{\ensuremath{n_\mathrm{H}}}
\newcommand{\pcmsq}{\ensuremath{\mathrm{cm}^{-2}}}
\newcommand{\chandra}{\textit{Chandra}}
\newcommand{\xmmn}{\textit{XMM-Newton}}
\newcommand{\cxo}{\textit{Chandra X-ray Observatory}}
\newcommand{\lae}{\lower 2pt \hbox{$\, \buildrel {\scriptstyle <}\over {\scriptstyle \sim}\,$}}
\newcommand{\gae}{\lower 2pt \hbox{$\, \buildrel {\scriptstyle >}\over {\scriptstyle \sim}\,$}}
\newcommand{\apropa}{\lower 2pt \hbox{$\, \buildrel {\scriptstyle \propto}\over {\scriptstyle \sim}\,$}}

\received{2018 February 09}
\revised{2018 April 26}
\accepted{2018 May 01}
\submitjournal{ApJ Letters}

\shorttitle{GW170817 Most Likely Made a Black Hole} 
\shortauthors{Pooley, Kumar, Wheeler \& Grossan}

\begin{document}

\title{GW170817 Most Likely Made a Black Hole}

\author{David Pooley}
\affiliation{Department of Physics and Astronomy, Trinity University, San Antonio, Texas}
\affiliation{Eureka Scientific, Inc.}

\author{Pawan Kumar}
\affiliation{Department of Astronomy, University of Texas at Austin, Austin, Texas}

\author{J.\ Craig Wheeler}
\affiliation{Department of Astronomy, University of Texas at Austin, Austin, Texas}

\author{Bruce Grossan}
\affiliation{Space Sciences Laboratory, University of California, Berkeley, CA}
\affiliation{Energetic Cosmos Laboratory, Nazarbayev University, Astana, Kazakhstan}

\correspondingauthor{David Pooley}
\email{dpooley@trinity.edu}

\begin{abstract}
There are two outstanding issues regarding the neutron-star merger event GW170817: the nature of the compact remnant and the interstellar shock. The mass of the remnant of GW170817, $\sim 2.7$ \msun, implies the remnant could be either a massive, rotating, neutron star, or a black hole. We report \chandra\ Director's Discretionary Time observations made in 2017 December and 2018 January, and we reanalyze earlier observations from 2017 August and 2017 September, in order to address these unresolved issues. We estimate the X-ray flux from a neutron star remnant and compare that to the measured X-ray flux. If we assume that the spin-down luminosity of any putative neutron star is converted to pulsar wind nebula X-ray emission in the 0.5--8 keV band with an efficiency of $10^{-3}$, for a dipole magnetic field with $3\times10^{11}$ G $<$ B $< 10^{14}$ G, a rising X-ray signal would result and would be brighter than that observed by day 107; we therefore conclude that the remnant of GW170817 is most likely a black hole.  Independent of any assumptions of X-ray efficiency, however,  if the remnant is a rapidly-rotating, magnetized, neutron star, the total energy in the external shock should rise by a factor $\sim$$10^2$ (to $\sim$$10^{52}$ erg) after a few years; therefore, \chandra\ observations over the next year or two that do not show substantial brightening will rule out such a remnant. The same observations can distinguish between two different models for the relativistic outflow, either an angular or radially varying structure. 


\end{abstract}


\section{Introduction}
The discovery of gravitational waves from a binary neutron star merger by LIGO \citep{2017PhRvL.119p1101A}, the associated electromagnetic signal that arrived with a delay of 1.74s \citep{2017ApJ...848L..12A}, and the precise localization \citep{2017Sci...358.1556C} have opened a new and exciting frontier that should provide answers to long-standing questions regarding  the synthesis of r-process elements \citep[e.g.][]{2017ApJ...848L..27T,2017Natur.551...80K} and the generation of relativistic jets and $\gamma$-ray photons \citep{2017Sci...358.1559K} in these events and possibly other high energy sources.  X-ray observations with the \cxo\ initially did not detect the event at 2--3 days post merger, but subsequent observations showed a brightening X-ray source 9 days after the merger \citep{2017ApJ...848L..25H, 2017ApJ...848L..20M, 2017Natur.551...71T}.

Of considerable interest is the nature of the compact remnant left behind by this event. Some theoretical works have suggested that the compact remnant is unlikely to be a neutron star that lived longer than a few months \citep{2017ApJ...851L..21V, 2017ApJ...850L..19M}. Others do allow a neutron star remnant with a relatively modest magnetic field to be consistent with early X-ray measurements \citep[][but see \S3.2 and \S4]{2017arXiv171101898Y}. Observations published to date are inconclusive. Here, we address anew the final product of the merger, and if and how observations could rule out or confirm either a neutron star or black hole compact remnant. We also discuss the unresolved issue of the nature of the propagating external shock driven by the merger. We requested Director's Discretionary Time \chandra\ observations in an attempt to address these issues.  

\begin{figure*}[htb!]
\includegraphics[width=\textwidth]{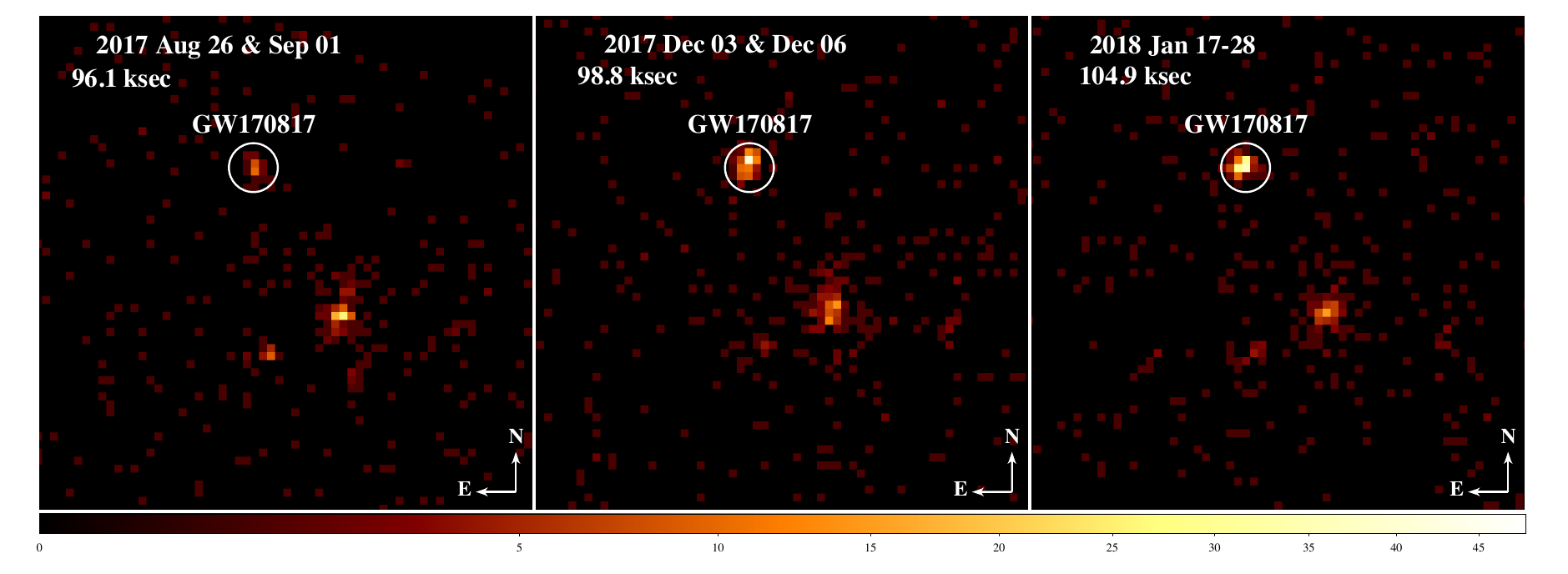}
\caption{\chandra\ images of the field of GW170817.  Each image is $30\arcsec \times 30\arcsec$ and made with counts in the range 0.5--8 keV.  The left image shows the combined data from Aug 26 and Sep 01 data \citep[presented in][and reanalyzed and discussed here]{2017Natur.551...71T}. The middle image shows the combined data from Dec 03 and Dec 06 presented and discussed here. The right image shows the combined data from Jan 17--28 presented and discussed here. The colorbar indicates the number of counts in each pixel. \label{fig:xim}}
\end{figure*}

The publicly available \chandra\ observations are described in \S2. In section \S3 we discuss the interpretation of the X-ray data. The main conclusions are summarized in \S4.

\section{Chandra observation and results}

We list in Table~\ref{tab:obs} the starting dates of all publicly available observations, their ObsIDs, and their exposure times in ks. All observations were taken with the telescope aimpoint on the Advanced CCD Imaging Spectrometer (ACIS) S3 chip.  Data reduction was performed with the chandra\_repro script, part of the Chandra Interactive Analysis of Observations (CIAO) software.  We used CIAO version 4.9 and calibration database (CALDB) version 4.7.6.  Images of the field of GW170817 are shown in Figure~\ref{fig:xim}.

\begin{deluxetable}{lrrrrl}[htb!]
\tablecaption{\chandra\ Observations of GW 170817 \label{tab:obs}}
\tablehead{
\colhead{Start Date} & \colhead{ObsID} & \colhead{Days After} & \colhead{Exp.} & \colhead{Net Counts} & \colhead{Count Rate}\\ 
& & \colhead{Merger\tablenotemark{a}}&  \colhead{(ks)} & \colhead{0.5--8 keV} &\colhead{cts ks$^{-1}$} 
}
\startdata
2017 Aug 26.44 & 19294 &   9.20 & 49.4 & 13.4\err{4.1}{3.4} & 0.27\err{0.08}{0.07} \\ \hline
2017 Sep 01.64 & 20728 &  15.38 & 46.7 & 18.4\err{4.7}{4.0} & 0.39\err{0.10}{0.09} \\ \hline
2017 Dec 03.07 & 20860 & 107.97 & 74.1 & 105.1\err{10.2}{10.2} & 1.42\err{0.14}{0.13} \\
2017 Dec 06.45 & 20861 & 111.07 & 24.7 & 34.7\err{6.2}{5.6} &1.40\err{0.25}{0.23} \\
\multicolumn{2}{l}{2017 Dec 03/06 joint}&109.38&98.8 & 139.8\err{11.8}{11.8} & 1.41\err{0.09}{0.17}\\ \hline
2018 Jan 17.89 & 20936 & 153.55 & 31.7 & 56.6\err{7.9}{7.2} & 1.78\err{0.25}{0.23} \\
2018 Jan 21.56 & 20938 & 157.13 & 15.9 & 27.8\err{5.6}{5.0} & 1.75\err{0.35}{0.31} \\
2018 Jan 23.32 & 20937 & 158.92 & 20.8 & 28.7\err{5.7}{5.0} & 1.38\err{0.27}{0.24} \\
2018 Jan 24.36 & 20939 & 159.94 & 22.2 & 23.7\err{5.2}{4.6} & 1.06\err{0.23}{0.21} \\
2018 Jan 28.18 & 20945 & 163.73 & 14.2 & 15.8\err{4.3}{3.7} & 1.11\err{0.31}{0.26} \\
\multicolumn{2}{l}{2018 Jan 17--28 joint}&158.59 & 104.9 & 152.6\err{12.4}{12.4} & 1.45\err{0.12}{0.12} \\ \hline
\enddata
\tablenotetext{a}{At the midpoint of the observation(s) being analyzed.}
\tablecomments{Horizontal rules separate epochs.  The last lines in the Dec and Jan epochs give the relevant information when considering all of the observations from the epoch as a single data point.}
\end{deluxetable}

In each observation, we extracted spectra from a 1\farcs0 radius region centered on GW 170817 and a 1\arcmin\ source-free background region to the northwest.  The net counts and count rate are given in Table~\ref{tab:obs}.

We performed a simultaneous fit of all nine unbinned source spectra in the 0.5--8 keV band with Sherpa using the modified \citet{1979ApJ...228..939C} statistic cstat and the simplex optimization method.  We fit the data with a power-law model with two absorption components.  We use the Tuebingen-Boulder ISM absorption model \citep{2000ApJ...542..914W} and fix one absorbing column to the Galactic value of $n_\mathrm{H} = 7.20 \times 10^{20} \textrm{cm}^{-2}$ calculated from the  Effelsberg-Bonn HI Survey \citep{2016A&A...585A..41W} using the online tool  at the Argelander-Institut f\"{u}r Astronomie\footnote{https://www.astro.uni-bonn.de/hisurvey/AllSky\_profiles/}.  We let the column density of the other absorption component vary to allow for absorption in the host galaxy and any absorption local to the event. 

In our fits, we require the same absorbing column for all spectra but allow the nine power-law indices and power-law normalizations to vary independently.  In addition, we perform another fit with both of the December observations required to have the same power-law index and normalization and all of the January observations required to have the same power-law index and normalization.  This resulted in a slightly different best-fit value for the additional column density. The results are given in Table~\ref{tab:fits}, separated by a horizontal rule for the two sets of simultaneous fits. All uncertainties are 1$\sigma$-equivalent (68\%) confidence intervals.  The reported fluxes are integrated from the unabsorbed models.  Uncertainties on those fluxes are calculated as the 68\%-confidence bounds of the integrated, unabsorbed fluxes from Monte Carlo realizations (1000 samples) of the best-fit models, taking into account the uncertainties in the best-fit parameters (using the sample\_flux command in Sherpa).

\begin{deluxetable*}{rlll|lll}[htb!]
\tablecaption{Power-law Fits to \chandra\ Spectra \label{tab:fits}}
\tablehead{
\colhead{Days} & \multicolumn{3}{c}{allowing for additional \nH\tablenotemark{b}} & \multicolumn{3}{c}{with no additional \nH} \\
\colhead{After}& \colhead{PL Index\tablenotemark{c}} & \colhead{Unabs.\ $F_\mathrm{0.5-8\,keV}$} & \colhead{cstat / dof} & \colhead{PL Index\tablenotemark{c}} & \colhead{Unabs.\ $F_\mathrm{0.5-8\,keV}$} & \colhead{cstat / dof} \\
\colhead{Merger\tablenotemark{a}}& & \colhead{($\ergcms$)} & & & \colhead{($\ergcms$)} 
}
\startdata
  9.20 & 1.1\err{0.4}{0.7} & 7.8\err{3.7}{2.8}$\times 10^{-15}$ & 1292.1 / 4607 & 0.7\err{0.5}{0.3} & 7.7\err{2.9}{2.2}$\times 10^{-15}$ & 1294.9 / 4608\\ 
 15.38 & 1.4\err{0.6}{0.3} & 7.9\err{7.0}{3.6}$\times 10^{-15}$ &               & 1.4\err{0.4}{0.4} & 7.4\err{2.6}{1.6}$\times 10^{-15}$ &              \\ 
107.97 & 1.9\err{0.1}{0.1} & 3.5\err{0.7}{0.6}$\times 10^{-14}$ &               & 1.4\err{0.1}{0.1} & 3.0\err{0.3}{0.3}$\times 10^{-14}$ &              \\
111.07 & 2.2\err{0.5}{0.3} & 3.3\err{1.2}{0.9}$\times 10^{-14}$ &               & 1.9\err{0.3}{0.2} & 2.6\err{0.6}{0.5}$\times 10^{-14}$ &              \\
153.55 & 1.7\err{0.2}{0.3} & 3.2\err{1.1}{0.8}$\times 10^{-14}$ &               & 1.3\err{0.2}{0.1} & 2.8\err{0.5}{0.4}$\times 10^{-14}$ &              \\
157.13 & 2.2\err{0.3}{0.4} & 3.2\err{1.3}{1.0}$\times 10^{-14}$ &               & 1.7\err{0.3}{0.2} & 2.6\err{0.5}{0.5}$\times 10^{-14}$ &              \\
158.92 & 2.2\err{0.4}{0.3} & 2.5\err{0.7}{0.5}$\times 10^{-14}$ &               & 1.6\err{0.2}{0.3} & 2.0\err{0.5}{0.4}$\times 10^{-14}$ &              \\
159.94 & 2.2\err{0.4}{0.3} & 1.8\err{0.9}{0.6}$\times 10^{-14}$ &               & 1.8\err{0.3}{0.3} & 1.5\err{0.4}{0.3}$\times 10^{-14}$ &              \\
163.73 & 2.5\err{0.5}{0.5} & 2.2\err{1.0}{0.9}$\times 10^{-14}$ &               & 1.9\err{0.4}{0.3} & 1.5\err{0.5}{0.4}$\times 10^{-14}$ &              \\ \hline
  9.20 & 0.8\err{0.6}{0.5} & 7.7\err{2.7}{2.1}$\times 10^{-15}$ & 1303.1 / 4617 & 0.7\err{0.5}{0.3} & 7.7\err{2.9}{2.2}$\times 10^{-15}$ & 1305.3 / 4618\\ 
 15.38 & 1.6\err{0.5}{0.5} & 7.8\err{2.4}{1.7}$\times 10^{-15}$ &               & 1.4\err{0.4}{0.4} & 7.4\err{2.6}{1.6}$\times 10^{-15}$ &\\ 
109.38 & 1.6\err{0.1}{0.1} & 3.0\err{0.2}{0.3}$\times 10^{-14}$ &               & 1.5\err{0.1}{0.1} & 2.9\err{0.3}{0.3}$\times 10^{-14}$ &\\ 
158.59 & 1.7\err{0.2}{0.1} & 2.2\err{0.2}{0.2}$\times 10^{-14}$ &               & 1.6\err{0.1}{0.1} & 2.1\err{0.2}{0.2}$\times 10^{-14}$ &\\
\enddata
\tablenotetext{a}{At the midpoint of the observation(s) being fit.}
\tablenotetext{b}{For the top set of fits, the best-fit additional column is $\nH = \ee{2.6\err{0.7}{0.8}}{21}\pcmsq$ and, for the bottom set of fits, is $\nH = \ee{1.8\err{0.5}{0.5}}{21}\pcmsq$.}
\tablenotetext{c}{PL index is the parameter $\beta$, where the number counts spectrum, $dN/dE/dA/dt \propto E^{-\beta}$.}
\tablecomments{The fits on the left allow for additional absorption beyond that through the Milky Way, while the fits on the right do not.  The top sets of fits allow different power-law normalizations and slopes for each \chandra\ observation listed in Table~\ref{tab:obs}.  The bottom sets of fits constrain the two observations of the December epoch (days 108 and 111) to have the same slope and normalization and the five observations of the January epoch (between days 153 and 164) to have the same slope and normalization.}
\end{deluxetable*}

We also perform the fits with no additional absorption component.  Compared to the fits allowing for additional absorption, these fits have a slightly worse fit statistic, slightly lower best-fit power-law indices, and roughly the same unabsorbed fluxes (Table~\ref{tab:fits}).

The large uncertainties on the power-law indices make it difficult to determine whether the X-ray spectrum has changed slope. To address this, we perform pairwise Kolmogorov-Smirnov tests on the detected photon energies and find marginal evidence for spectral evolution.  The K-S tests show probabilities of the detected photon energies being drawn from the same parent distribution in the $\sim$10\%--100\% range, which indicates consistency of the observed spectra.  The smallest probability for being drawn from the same parent distribution is for the first and fourth observations (ObsIDs 19294 and 20861), which have a 6\% probability of being drawn from the same parent distribution; this is not convincingly small enough to claim a spectral change.

\begin{figure}[htb!]
\includegraphics[width=\columnwidth]{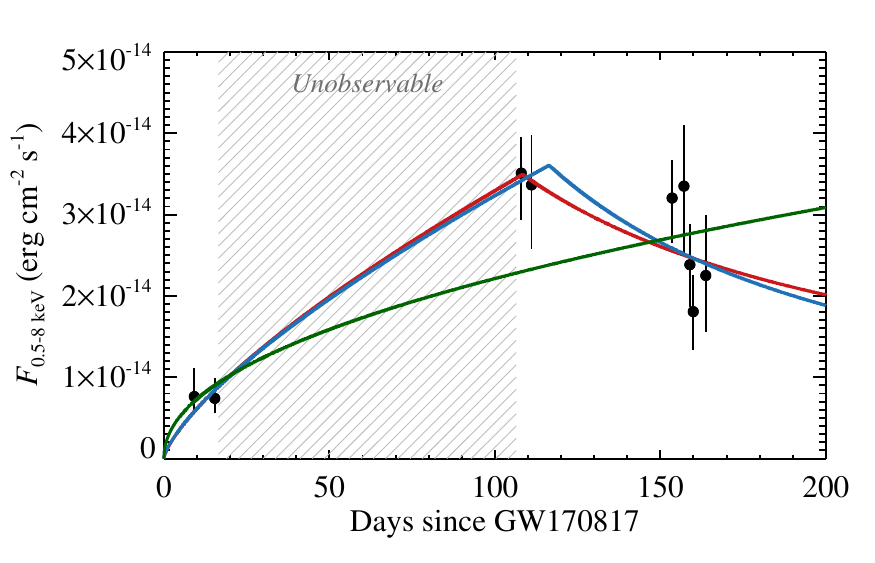}
\caption{\chandra\ light curve of the unabsorbed 0.5--8 keV flux from the available observations of GW170817.  Uncertainties in flux take account of uncertainties in the spectral fits; see text for details.  The hatched region indicates when GW170817 was behind the Sun and unobservable with \chandra.  The green line indicates the best fit single power-law $\Fx \sim t^{a_1}$ to the data, with $a_1 = 0.5$. The red and blue lines show the best fit broken power laws, $\Fx \sim t^{a_1}$ before $t_\mathrm{break}$ and  $\Fx \sim t^{a_2}$ after.  The red line shows the best fit when $a_2$ is fixed at $-0.9$ (expected when there is no energy being added to the forward shock and the shock front is moving at a relativistic speed), and the blue line shows the best fit when $a_2$ is fixed at $-1.2$ (expected for a sub-relativistic shock); in both cases, the best-fit $a_1$ is $0.7$.    \label{fig:xlc}}   
\end{figure}

Although there is a considerable amount of stochastic variation in the calculated fluxes (and count-rates), there does seem to be an indication that the temporal behavior at days 153--163 is qualitatively different than at earlier times. If we take the two December observations together as a single flux point and the five January observations together as a single flux point, we fit the four-point X-ray light curve with a single power-law model in time, $\Fx \sim t^{a_1}$, and find that $a_1 = 0.5$ with a $\chi^2$ of 7.8 for 2 degrees of freedom.  If we use all flux measurements separately in the fit, we find $a_1 = 0.5$ with a $\chi^2$ of 15.6 for 7 degrees of freedom.  Neither of these fits is acceptable ($p$-values of 0.02 and 0.03, respectively), but we plot a power law with index 0.5 as the green line in Figure~\ref{fig:xlc} for reference.

\begin{figure}[t!]
\includegraphics[width=\columnwidth]{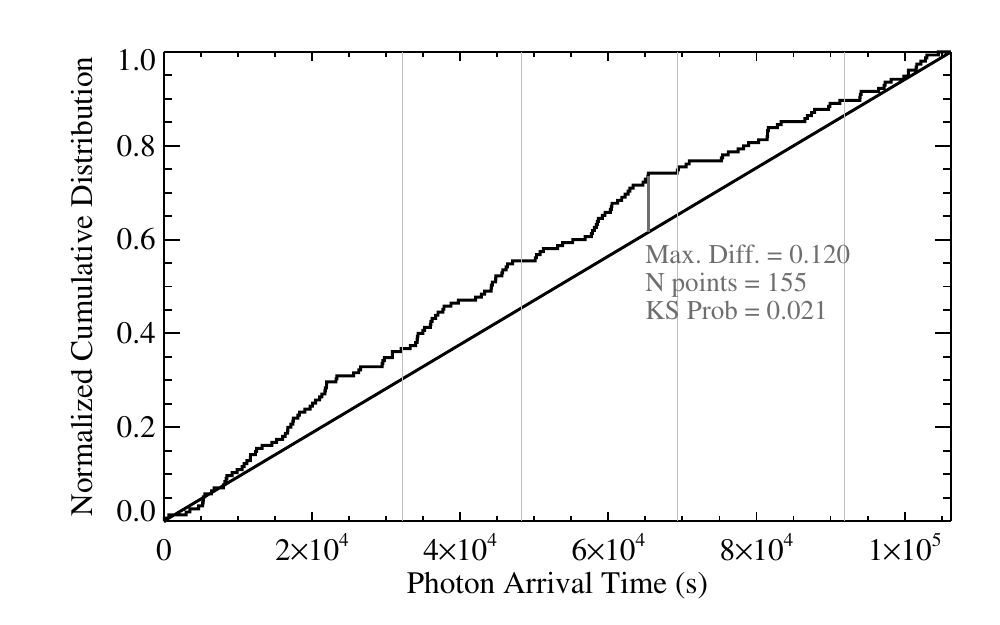}
\caption{Cumulative distribution of the photon arrival times from the January observations concatenated together, with faint gray lines marking the divisions between observations.  This is compared to a uniform distribution with a Kolmogorov-Smirnov test.\label{fig:janks}}
\end{figure}

The likely reason for the poorness of these fits is the possible fading trend seen in the January observations, as suggested by \citet{2018arXiv180106164D}.  In the most general case, we nonparametrically test whether the January observations are consistent with a constant count rate using a K-S test. The photon arrival times (during ``good-time intervals'') were compared to a uniform distribution. The five individual observations were concatenated into one, with the next observation beginning where the previous left off.  The K-S probability that the photon arrival times are drawn from a uniform distribution is only 2.1\%.  Figure~\ref{fig:janks} shows the cumulative distributions with faint gray lines marking the observation divisions. The maximum difference occurs near the division between the third and fourth January observations.

Given the poor fit with a single power law and the suggestive evidence for fading from the K-S test, we next fit a broken power-law model, $\Fx \sim t^{a_1}$ before $t_\mathrm{break}$ and  $\Fx \sim t^{a_2}$ after $t_\mathrm{break}$. Because there is little data after a possible break to constrain the decay slope, we fix $a_2$ to one of two values: $-0.9$ if  the shock front is moving at relativistic speeds and $-1.2$ if the shock front is sub-relativistic.  For $a_2 = -0.9$, we find $a_1 = 0.7\err{0.1}{0.6}$ with $t_\mathrm{break} = 108\err{13}{43}$ days. For $a_2 = -1.2$, we find $a_1 = 0.7\err{0.1}{0.1}$ with $t_\mathrm{break} = 116\err{10}{29}$ days.  The two fits have $\chi^2$ values of 5.6 and 5.4, respectively, for 6 degrees of freedom.  We plot these best fits in Figure~\ref{fig:xlc}.      

\section{Interpretation of the Chandra data}

In \S3.1 we conclude, as have others \citep{2018Natur.554..207M,2017ApJ...848L..25H,2017ApJ...848L..20M,2017Natur.551...71T,2018ApJ...853L...4R}, that the X-ray data from about 9 to 111 days after the GW detection are consistent with an expanding shocked plasma emitting synchrotron radiation due to its interaction with the ISM, and with energy input to the plasma during this time. The possible fading seen in \chandra\ data from days 153--163 would be consistent with reduced or no energy input. In \S3.2, we show that considerations of X-ray optical depth give the time after which emission from a  recently formed magnetic neutron star or a pulsar wind nebula (PWN) would dominate. The X-ray flux measured during days 107--111 and 153--163 is much smaller than that expected if a neutron star remnant were formed and lived for $\gtrsim 10^2$ days. This suggests that a black hole was born in this merger well before the observations on day 107. We propose other stringent tests of this conclusion in \S\ref{future}.

\subsection{External shock origin of X-ray light curve}
\label{shock}
The radio spectrum reported in \citet{2018Natur.554..207M} gave a relatively precise log slope of $-0.61\pm0.05$, i.e. $f_\nu\propto \nu^{-0.61\pm0.05}$, for data through day 107. Our best fit X-ray power-law index, though poorly constrained, is consistent within uncertainties with those in the radio. \cite{2018arXiv180103531} extend the single consistent power law conclusion to day 160.

The most conservative scenario, therefore, is that the radio and the X-ray photons are produced in the same source via the synchrotron radiation mechanism. The synchrotron cooling frequency in a relativistic shock wave can be shown to be \citep[e.g.,][]{2001ApJ...554..667P,2002ApJ...568..820G}
\begin{equation}
   h\nu_c \sim 4.5\,{\rm keV} \; E_{50}^{-1/2} n^{-1} \epsilon_{B,-4}^{-3/2}
           t_7^{-1/2} (1+Y)^{-2},
\end{equation}
where $E$ is the energy in the shocked plasma, $n$ is the number density of particles in the medium in the vicinity of the neutron star merger, $\epsilon_B$ is the fraction of the thermal energy of the shocked plasma in magnetic fields, $t$ is the time since the merger event in the observer frame, and $Y$ is the Compton $Y$-parameter, which is less than 1 based on radio and X-ray data; integer subscript for a variable $X$, i.e. $X_m$, is a convenient short hand notation for $X/10^m$ in cgs units. The density $n$ is expected to be of order $0.1$ cm$^{-3}$ or less in the interstellar medium of the host galaxy of GW170817. Thus, $h\nu_c > 10$ keV even a year after the merger. Our X-ray observations are below the expected cooling frequency.

 The observed synchrotron flux at 1 keV  is given by \citep[e.g.][]{2015PhR...561....1K}
\begin{equation}
   f_\nu \sim 70 \,{\rm nJy}\; E_{50}^{1.3} \epsilon_{e}^{1.2} n^{1/2} 
         \epsilon_{B,-4}^{0.8} t_7^{-0.9},
         \label{obsflux}
\end{equation}
where $\epsilon_{e}$ is the fraction of shock energy given to electrons. In deriving the above expression we have taken the electron energy distribution to be $dn(e)/de \propto e^{-p}$ with $p=2.2$ as suggested by the X-ray and radio spectra that give  $f_\nu\propto \nu^{-0.6}$.

For constant shock energy, equation (\ref{obsflux}) suggests that the X-ray flux should decline with time. Instead, the X-ray and radio light curves are rising with time from day 9 until at least day 111. For $f_\nu\propto t^\alpha$, we find from equation (\ref{obsflux}) that energy in the shocked matter is increasing with time as
\begin{equation}
   E \propto t^{(\alpha+0.9)/1.3}.
   \label{E-shock}
\end{equation}
The flux at 1 keV at $10^2$ days (1.4 nJy) is larger than the flux at 9 days by a factor of $\sim$5, implying that $\alpha\sim 0.7$. We thus conclude from equation (\ref{E-shock}) that the energy in the shocked plasma increased by a factor $\sim$20 between these two epochs. There are two ways that the energy in the external shock could increase with time. One is that the compact remnant left over in the merger continues to pump energy into the outward moving shock front. The other is that energy is supplied to the part of the external shock wave that we see by slower moving ejecta or a structured jet\footnote{A jet with energy per unit solid angle, $dE/d\Omega$, that is independent of angle $\theta<\theta_j$, and drops to zero for $\theta>\theta_j$ is referred to as a {\it top-hat jet}. A jet where $dE/d\Omega$ drops to zero as some smooth function of $\theta$ is called a structured jet.}.

The first possibility, continuous production of energy from the compact object, is unlikely to work. This is because there is $\sim$\ee{3}{-2} \msun\ of debris between the central compact remnant and the external shock front \citep[e.g.,][]{2017Sci...358.1559K}. If the central compact object produces {$\sim 10^{52}$ erg of energy, and a good fraction of this energy is used to accelerate the debris, then the speed of the debris would be $\sim 0.7$c. This speed is smaller than the outward velocity of the {relativistic} shock-front. Any such energy will not be added to the shock until the shock speed has finally fallen below $\sim 0.7$c. This is expected to take a year or more (\S\ref{remnant}). Observations carried out then would be able to constrain the total energy in the outflow, as discussed in \S\ref{future}. 

We are thus left with the second possibility: at 107 days, there is much more energy in the shocked medium within our cone of visibility, i.e., $\delta\theta \sim \gamma^{-1}$, than there was at 9 days, where $\gamma$ is the Lorentz factor of the shocked fluid within $\delta\theta$. There are two possible ways this could be true: slower moving ejecta caught up with the external shock or we see more of the shock as it decelerates and beaming effects decrease.

We first consider the scenario where the unexpected shock energy is supplied by slower-moving, perhaps quasi-spherical, ejecta that catches up with the decelerating shock front. Let us take the distribution of energy with ejecta velocity ($\beta\gamma$) such that

\begin{equation}
dE/d(\beta\gamma)\propto [\beta\gamma]^{-\eta_E}, \quad{\rm or}\quad
     E(>\beta\gamma) \propto [\beta\gamma]^{1-\eta_E},
   \label{dE-dgam}
\end{equation}
where the exponent $\eta_E$ is determined from the observed light curves. Combining the energy equation \citep[e.g.,][]{2002ApJ...568..820G}
\begin{equation}
   E\sim (4\pi/3) r^3 n m_p c^2 (\beta\gamma)^2,
\end{equation} 
with the relation between the shock front radius $r$ and observer time $t$
\begin{equation}
   t \sim \int {dr\over \beta c} (1-\beta)
\end{equation}
yields
\begin{equation}
   \beta\gamma \apropa \left(E/n\right)^{1/8} t^{-3/8}.
\label{betagamma}
\end{equation}
Using equation (\ref{dE-dgam}) we obtain
\begin{equation}
   E(>\beta\gamma) \apropa t^{-{3(1-\eta_E)\over 7+\eta_E}},
\end{equation}
which is consistent with equation (\ref{E-shock}) provided that
\begin{equation}
   {3(\eta_E-1)\over 7+\eta_E} = {\alpha + 0.9 \over 1.3}\quad {\rm or}
    \quad \eta_E = {10.2 + 7\alpha\over 3 - \alpha}.
\end{equation}
The X-ray data give $\alpha\sim 0.7$, and therefore the rising light curve can be explained if $dE/d(\beta\gamma)\apropa[\beta\gamma]^{-6.6}$,  which is roughly consistent with the conclusion of \citet{2018Natur.554..207M}.

A rising light curve can also occur when $E$ is a rapidly decreasing function of angle $\theta$ measured from the jet axis and the observer is located off axis, as was pointed out by \citet{2017arXiv171203237L} for GW170817. In the case of either a top-hat structure or a more gently-shaped angular structure, an off-axis observer will see a rising light curve as the external shock decelerates and the Lorentz beaming decreases. The rate of increase for a top hat jet scales as $t^3$, which is inconsistent with the data
\citep{2018Natur.554..207M, 2018ApJ...853L...4R}.  For a structured jet, the rate of rise depends on how the energy declines with angle.   In this case, an off-axis observer finds the energy in the shocked medium to increase with time because photons are detected from a larger fraction of the external shock. It can be shown that the rising X-ray flux between $\sim$10 and 10$^2$ days ($f_x \propto t^{0.7}$) can arise from a structured jet when $dE/d\theta\apropa (1 + \theta/\theta_c)^{-4}$, ($\theta_c$ is the angular size of the jet core) with $\theta_c\sim 3^\circ$, and the observer is located off axis at an angle $\sim 30^\circ$ with respect to the jet axis.

The data for GW170817 are consistent with both of these possibilities: slower moving ejecta that caught up with the external shock or beaming effects from a structured jet. This degeneracy may be 
removed by careful modeling of afterglow light curve data about a year after the merger.  If the rise is due to a structured jet with angular size of the core $\sim 3^\circ$, then the X-flux should stop increasing at $t\sim10^2$ days; there is evidence for this flattening in the \chandra\ data at around day 160 (Figure~\ref{fig:xlc}). 

We note that the total energy in the shocked ISM is highly uncertain because of the unknown parameters $n$, $\epsilon_B$ and $\epsilon_e$. We return to the discussion of the nature of the external shock in \S\ref{future}.

\subsection{Merged remnant a black hole or a neutron star?}
\label{remnant}
\medskip

The merger of two neutron stars with mass 1.48$\pm0.12$M$_\odot$ and 1.26$\pm0.1$M$_\odot$ --- where the merged object has a mass of  2.74$^{+0.04}_{-0.01}$M$_\odot$ \citep{2017PhRvL.119p1101A} --- could result in either a massive neutron star or a black hole. There might also be a debris disk that gets accreted onto the central object over a period of time. The observed X-ray luminosity at around 10$^2$ days is much larger than the expected value from any debris disk, so we do not consider such a source further here. 

The large orbital angular momentum of the inspiralling neutron stars make the rapid rotation of the compact remnant inevitable. If the surviving compact remnant is a neutron star, then the expectation is that dynamo processes will work rapidly in the differential rotation of the remnant to produce substantial magnetic fields, perhaps of magnetar strength ($\sim  10^{14}$ G) \citep{1992ApJ...392L...9D,2003ApJ...584..954A,2014MNRAS...439..3916}. Having been born in the highly energetic, shearing environment of this binary system, a neutron star with less than $10^{13}$ G is highly unlikely. We assume substantial initial rotation and magnetic field in the subsequent discussion unless the context explicitly calls for other values. 

The bolometric spin-down luminosity of a neutron star due to dipole radiation is \citep[e.g.,][]{1983bhwd.book.....S}
\begin{equation}
L_{d}(t)\approx(6\times10^{45}{\rm erg\,s}^{-1})B_{13}^2\,P_{-3}^{-4}\,
\left(1+t/t_{_{\rm SD}}\right)^{-2}\,,
  \label{sd-power}
\end{equation}
where
\begin{equation}
t_{_{\rm SD}}\approx(5\times10^6\,{\rm s})\,B_{13}^{-2}\,P_{-3}^{2}\,,
  \label{ns-sd-power}
\end{equation}
is the spin-down time, $B_{13}$ is the dipole magnetic field in units of $10^{13}$ G, and $P_{-3}$ is the rotation period in units of milliseconds. 

The rotational energy of a millisecond period massive neutron star is $\sim$\ee{3}{52}~erg. From equation (\ref{ns-sd-power}), this energy is deposited in the neutron star merger debris on a timescale less than $10^2$ days provided that the dipole magnetic field strength is larger than \ee{7}{12}~G. A fraction of the spin-down energy goes into heating the debris, and a fraction is used to accelerate the ejecta of mass, $\sim$\ee{3}{-2} \msun, to speed $\beta\Gamma\sim 1.1$. From equation \ref{betagamma}, the speed of the external shock decreases with time as
\begin{equation}
   \beta\gamma \sim 1.3\, t_7^{-3/8} (E_{50}/n_{-2})^{1/8},
  \label{beta-gamma}
\end{equation}
so the debris moving with $\beta\gamma\sim 1.1$ will catch up with the decelerating ISM shock front in about 2 years and transfer a substantial fraction of its energy to the external shock. 

Consideration must also be given to the fraction of the unobscured spin-down power that could appear directly in the 0.5--8 keV X-ray band. \citet{2014MNRAS...439..3916}, and \citet{2016ApJ...819..15p} provided estimates for optical, UV and X-ray luminosities from the dipole wind of a neutron star remnant and its interaction with the material ejected in the merger. These authors considered a wide variety of physical processes such as the power law spectrum for spin-down radiation, adiabatic losses, pair production, and absorption and scattering of photons as they travel through the system, which is considered to be spherically symmetric. According to \citet{2014MNRAS...439..3916}, the X-ray luminosity should peak between 1 and 10 days after the merger, depending on the mass of the debris and the magnetic field strength of the remnant. They argued that the ejecta would be fully ionized and therefore of low opacity, giving a measured X-ray luminosity roughly between 10$^{-1}$ and 10$^{-3}$ of the spin-down power (depending on various parameters), declining as $t^{-2}$ for $t\gae 10$ days. The resulting X-ray flux would then considerably exceed that observed for GW170817 in this early epoch. 

The IR, optical and UV light curves and spectra for this event, however, showed a photospheric temperature of $\sim2\times10^3$~K and many absorption lines in the UV about 10 days after the merger (e.g. Cowperthwaite et al. 2017). This suggests that the ionization state of the ejecta was not very high. This might be because the fraction of spin-down power deposited in the ejecta was smaller than estimated by \citet{2014MNRAS...439..3916}. Here we take a more conservative approach and adopt the fraction of spin-down power converted to X-ray luminosity from the extensive data available for pulsar wind nebulae (PWN). Our estimates of the X-ray flux are smaller than earlier works, and hence the conclusions we draw concerning the possible existence of a neutron star remnant are more robust. 

We use the data for pulsars in our Galaxy to estimate the X-ray flux that would be produced assuming that the remnant of GW170817 is a long-lived massive neutron star.  We show below that the \chandra\ flux at 107 days is smaller than expected for a pulsar remnant even for our conservative estimates.

The bolometric EM luminosity of the {\it pulsed} radiation from the Crab is about 0.1\% of the spin-down luminosity. Of this flux, only about 10\% comes out as 1--10 keV photons \citep{2011ApJ...743...38D, 2014RPPh...77f6901B}. Thus, the fraction of the spin-down rate of energy loss released as pulsed X-ray photons is $\sim10^{-4}$ \citep{2011ApJ...743...38D}. The fraction of spin-down energy going into pulsed X-ray photons does not seem to be dependent on pulsar age or surface magnetic field strength \citep[see Table 8 of][]{2011ApJ...743...38D}.  Geminga, at $3\times10^5$ years, has an X-ray efficiency within a factor 3 of the Crab, which is only 10$^3$ yrs old. Therefore, a crude estimate is that $\sim10^{-4}$ of the spin-down luminosity of the merged neutron star in GW170817 should be expected as 1--10 keV {\it pulsed} X-rays\footnote{ Note that, as pulsed X-ray radiation might be beamed in a cone of few 10s of degrees opening angle, an observer would not directly measure this pulsed component unless their line of sight intersected this small cone. }. The total X-ray luminosity emitted directly by the NS} might be a factor of 10 larger \citep{1996A&A...311L...9V}. 

A larger fraction of the pulsar spin-down power is converted to EM radiation in the pulsar-wind nebula (PWN) which is produced when the pulsar wind  interacts with the surrounding medium. A fraction $\sim$10$^{-3}$ of the spin-down power is converted to 0.5-8 keV X-rays in the PWN \citep{2008AIPC..983..171K}. The fraction of spin-down energy released in the X-ray band in the PWN is larger for younger pulsars and pulsars with higher rotation speed, although the scatter is large in these correlations \citep{2008AIPC..983..171K}.

A fairly conservative estimate for the 0.5--8 keV X-ray luminosity of a neutron star star remnant of GW170817, assuming a millisecond rotation period, is thus $\sim$10$^{-3}$ of the spin-down power. Using equation (\ref{sd-power}), we obtain the X-ray flux from the PWN associated with any long-lived neutron star remnant of GW170817 to be
\begin{equation}
f_x \sim \left\{
\begin{array}{ll}
    \hskip -7pt ~3\times10^{-11}\, {\rm erg\,cm^{-2}\,s^{-1}}\, B_{13}^2\,P_{-3}^{-4} \,  & B_{13} < 2.2 t_{6}^{-1/2} P_{-3} \\ \\
 \hskip -7pt ~7\times10^{-10}\, {\rm erg\,cm^{-2}\,s^{-1}}\,B_{13}^{-2}\,t_{6}^{-2} \, & B_{13} > 2.2 t_{6}^{-1/2} P_{-3}
\end{array}
\right.
  \label{fx-pwn}
\end{equation}
where we took the distance to the object to be 40 Mpc \citep{2017ApJ...848L..12A}. This is to be compared to the observed, unabsorbed flux of $f_x \sim 8\times 10^{-15}\,\ergcms$ at 10 days and $\sim 3\times 10^{-14}\,\ergcms$ at 107 days.

In order to determine the constraints on the properties of a possible neutron star remnant, it is important to know when the ejecta are optically thick to X-rays. The ejecta in a binary neutron star merger event consist of nuclei of high atomic number that have large cross-sections for absorbing X-ray photons via the bound-free process. The X-ray opacity for binary neutron star merger ejecta can be found in, e.g., \citet{2016MNRAS.459...35H}. At 5 keV, the mass absorption coefficient due to elements with atomic number between 40 and 92 is estimated to be $\sim$\ee{5}{2} cm$^2$ g$^{-1}$, and at higher photon energies it declines approximately as $E^{-1.8}$. The optical depth of the ejecta moving with speed $v_\mathrm{ej}$, after time $t$ is $\tau_\mathrm{5keV} \sim 8\, M_\mathrm{ej,-2} t_6^{-2} v_\mathrm{ej,10}^{-2}$; where $M_\mathrm{ej,-2} = M_\odot/10^2$ is the isotropic equivalent mass of matter ejected along our line of sight, and $v_\mathrm{ej,10} = v_\mathrm{ej}/10^{10}$~cm~s$^{-1}$. The X-ray photons are strongly absorbed for the earliest observations at $\sim 10$ days, but not for $t \gtrsim 10^2$ days. 

At $t=10$ days, the unabsorbed X-ray flux from a putative neutron star would be larger than the observed flux by a factor $\gtrsim 10^3$ for $B\lesssim 10^{14}$ G, but opacity effects must be considered at this epoch. If $M_\mathrm{ej,-2}>1$ as observations suggest \citep[e.g.,][]{2017Sci...358.1559K,2017Natur.551...80K}, at $t=10$ days, $\tau_\mathrm{5keV}>10$. Thus, the luminosity of X-ray photons produced in the PWN at 10 days that can escape is smaller than the unabsorbed luminosity by a factor $\sim$10$^5$. According to equation \ref{fx-pwn}, the neutron star flux is maximized for $B_{13} \sim 2.2$, and for a millisecond period, would be $f_x \sim 1.5\times10^{-10}\ergcms$. This maximum flux then, for the given absorption, is still smaller than that observed. We therefore do not expect a neutron star to be observable at this early epoch.

At $t=107$ days when the ejecta are expected to be optically thin, our conservative estimates for the X-ray flux from a PWN would be a factor of a few larger than the observed flux for the typical magnetar magnetic field strength $\sim10^{14}$\,G. From equation \ref{fx-pwn}, assuming a millisecond period, the predicted flux is maximum for a field of $B_{13} \sim 0.7$, giving  $f_x \sim 1.5\times10^{-10}\,\ergcms$, a factor of $\sim 500$ times brighter than observed.  A neutron star would therefore be unobservable only for a field $B\lesssim 3\times10^{11}$ G or stronger than $B\sim 1.4\times10^{14}$ G.

The fact that the temporal and spectral behaviors of the radio and X-ray flux are consistent with a single emission component at 107 days effectively rules out any additional component with flux comparable to or even somewhat smaller than the observed value. The observed X-ray flux at $t=107$ days thus rules out the possibility that the remnant of GW170817 is a long-lived neutron star with $3\times10^{11}\lesssim B\lesssim 10^{14}$ G.
If the field exceeds $10^{14}$ G, as may be expected, the spin-down energy would be deposited and thermalized very early while the remnant is still optically thick. This energy must, however, eventually be deposited in the external ISM shock, as we discuss in the next section.

\section{Future Light Curve Diagnostics for Remnant and Jet}
\label{future}
\label{Predictions}

Two important questions remain about GW170817: (1) what is the final remnant of the event, and (2) what is the nature of the relativistic jet? 

\subsection{Diagnostics for the Remnant of GW170817}

In \S\ref{remnant}, we made the case that GW170817 left behind a black hole because a new-born neutron star would be much brighter in X-rays at $\sim 10^2$ days than was observed. Rather, the X-ray emission through day 163 is fully consistent with an origin from a jet interacting with the local ISM. No additional source of X-ray emission (i.e., spin-down emission from a neutron star) is required.

Our estimate of the X-ray flux at $\sim 10^2$  days from a new-born neutron star was based on an observed relationship between pulsar spin-down luminosity and X-ray luminosity, which may not hold for this neutron star. Future X-ray observations can overcome this ambiguity. In \S\ref{remnant}, we noted that any spin-down energy input would not yet have affected radiation from the external shock at $\sim10^2$ days because it would not have had time to propagate to the decelerating shock front. If there is spin-down energy, however, it must eventually be deposited in the ISM shock, resulting in dramatic brightening. 

For $B > 3\times10^{12}$ G, the kinetic energy of rotation of the neutron star ($\sim 3\times10^{52}$ erg) is carried away by the magnetic dipole wind in one year or less. A fraction of this energy\footnote{For $B>10^{14}$G, the spin-down time is less than a day. As the ejecta is highly optically thick at this time, most of the radiation energy becomes bulk kinetic energy via adiabatic expansion. Since the system remains  optically thick to X-ray photons for several 10s of days,  much of the spin-down energy, even for a weaker magnetic field with a longer spin-down time, is eventually converted to bulk kinetic energy of the ejecta. Moreover, the bolometric radiative efficiency of a PWN (including in the well-studied Crab nebula) is a few percent, thus most of the spin-down energy should be in the bulk motion.} would go into accelerating the ejecta and then be deposited to the shock when the ejecta catch up to it. The predicted deposition is much larger than the total energy currently in the external shock ($\sim 10^{49}$--$10^{50}$ erg). When that deposition happens, the X-ray light curve would increase very roughly as $t^2$ during or after 2019, corresponding to a total increase of 1--2 orders of magnitude in $\sim$1--2 years. This scenario is represented by the associated rising curves in Figure~\ref{futureplot}. Because a neutron star with $B < 10^{12}$ G is very unlikely, if we do not see this dramatic brightening, we can conclude with certainty that the remnant is a black hole. 

\subsection{Diagnostics for the Nature of the Jet}

As described in \S\ref{shock}, there are two well-known models of the jet; one of these posits that the jet energy per unit solid angle and the Lorentz factor declines with angle from the jet axis, the ``structured jet" model of \citet{2017arXiv171203237L}. The other model invokes a jet with radial structure such that the jet speed decreases with increasing distance inward from the front surface of the jet, but with jet properties that do not vary with angle \citep{2017Sci...358.1559K,2017ApJ...834...28N}. 

Both of the proposed models successfully explain the radio and X-ray data  from day 9 to day 111 as synchrotron radiation from the ISM shock-heated by the relativistic outflow produced in the merger, with a rise in the light curve due to energy of the shocked ISM increasing between these times. The structured jet is predicted to reach maximum brightness within a few months of the day 107--111 observations; the radial jet will not peak until  $\sim$ day 530 (around 2019 Feb), with a 1/e fading time of $\sim$1 yr (\S\ref{shock}). Fading may already be present in the day 153--163 observations, but the data are not conclusive. Additional X-ray observations will distinguish between these models (Figure~\ref{futureplot}).

We may also learn about the microphysics of the shocked material if we  observe the epoch when the synchrotron cooling frequency passes through the X-ray band, thus breaking an important degeneracy in shock parameters. The observed flux scales as $n^{1/2} B^{1.6}$, and hence we cannot determine $n$ and $B$ separately from radio or X-ray flux alone. Measurement of the synchrotron cooling frequency can break this degeneracy between $n$ and $B$. The cooling frequency currently lies above the X-ray band of \chandra\ and \xmmn\ and decreases with time as $t^{-1/2}$. The cooling frequency could thus shift into the X-ray band during 2019 and would be seen as steepening of the spectral slope, i.e., a depression in the higher energy channel flux. 

\subsection{Future Observations}

\begin{figure*}[ht]
\centering
\includegraphics[width=.95\textwidth]{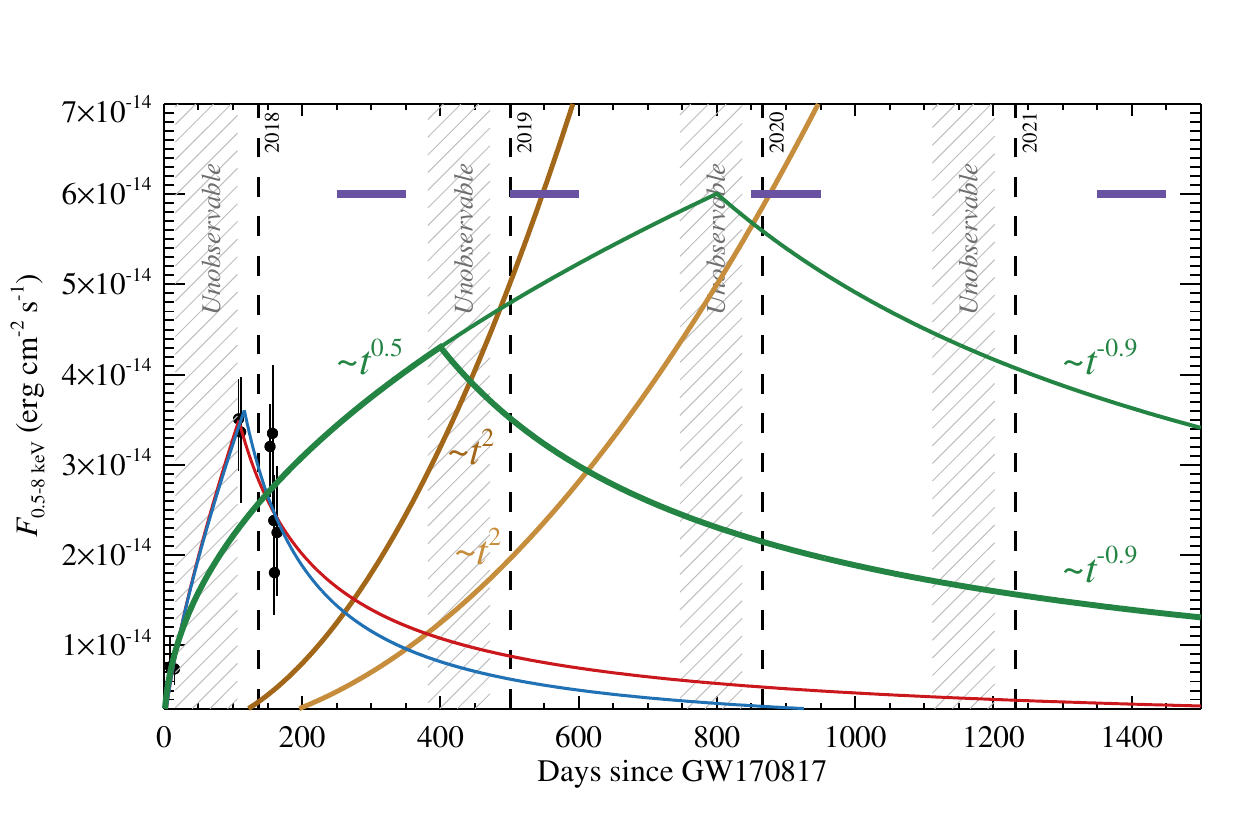}
\caption{An extension of Figure~\ref{fig:xlc} showing various possibilities for the long-term behavior of GW 170817. The red and blue curves show the same fading starting around day 109 as in Figure~\ref{fig:xlc}. Extrapolations of the early, rising X-ray flux with example jet breaks at either 400 or 800 days, and then subsequent power-law declines in flux are shown in green.  The brown curves show flux rising as $t^2$, which might be expected for PWN emission catching up with the external shock and depositing significant energy at 1--2 years after the merger. Given the potentially complicated sum of some form of a green and some form of a brown curve, the best way to disentangle their contributions is with a long-term, well-thought-out observation plan. A \chandra\ observation during each of the intervals indicated with horizontal purple lines would distinguish the possibilities.}  \label{futureplot}
\end{figure*}

Future X-ray observations could both test for the possible presence of rising PWN emission and also constrain the jet break time and rate of decline.  Because there is a reasonable chance that both of these could manifest in the next year, a long-term strategy is needed to fully understand the outcome of this merger.  In Figure~\ref{futureplot}, we indicate with purple lines four intervals during which sensitive X-ray observations would both test for the presence of rising emission from a PWN as well as distinguish between jet models.

\section{Conclusions}

The \chandra\ X-ray data for the neutron star merger event GW170817 obtained 107--111 days after the merger show that the rate of increase of the X-ray flux is essentially the same as the radio flux increase reported by \citet{2018Natur.554..207M}.  The extrapolation of the well-defined radio spectrum to the X-ray band is also consistent with the measured X-ray flux. These results suggest that radio and X-ray photons are produced in the same source, which is most likely the shocked interstellar medium (ISM). The rising X-ray flux between 9 and 111 days requires the observable energy in the shocked plasma to increase by a factor $\sim$20 during this period. That might be due to more energy resident in slower moving ejecta where the energy scales with shock-front velocity $\beta\gamma$ as $(\beta\gamma)^{-6.6}$, which is roughly consistent with radio data as reported by \citet{2018Natur.554..207M}, or due to a structured jet with $dE/d\theta\apropa (1 + \theta/\theta_c)^{-4}$ with $\theta_c\sim 3^o$ \citep{2017arXiv171203237L}.

The X-ray data at 107--111 days suggest that the remnant is not a neutron star with magnetic field $\gtrsim10^{12}$ G. This, in turn, suggests that the merged object is most likely a black hole. Additional checks of the possibility of a stable neutron star remnant could be provided by continued monitoring of the system in X-rays and radio. If the remnant is a stable neutron star with magnetic field $\gtrsim$10$^{12}$~G (so that the spin-down time is less than a few years), then in about a year the X-ray and radio fluxes from the external shock should increase by a large factor when the expanding PWN with energy $\sim$10$^{52}$~erg, and speed $\beta\gamma\sim 1.1$, catches up with the decelerating external shock (equation\ \ref{beta-gamma}).

Possible fading after 107--111 days (as may be indicated by the observations on days 153--163) may be a result of a structured jet and is still consistent with our arguments for a black hole remnant.

\acknowledgements
We thank Belinda Wilkes and \chandra\ team for carrying out the \chandra\ Director's Discretionary Time observations and E.\ L.\ Robinson, U. Nakar, and C.\ Froning for very useful discussions. JCW is supported in part by the Samuel T.\ and Fern Yanagisawa Regents Professorship. 

\facility{CXO}

\software{CIAO \citep{2006SPIE.6270E..1VF}, Sherpa \citep{2001SPIE.4477...76F}}

\end{document}